\documentclass[a4paper,11pt]{article}
\pdfoutput=1 

\usepackage{jinstpub}

\graphicspath{{fig/}}

\title{\boldmath Design of a $\mu$SR spectrometer with ultrahigh arrays at China Spallation Neutron Source}

\author[a]{J. Y. Dong,}
\author[a,1]{Z. W. Pan,\note{Corresponding author.}}
\author[a]{T. Y. Yang,}
\author[a]{Z. B. Lin,}
\author[a]{Z. Wang,}
\author[a]{Z. Y. He,}
\author[a]{Z. Chen,}
\author[b,c]{H. T. Jing,}
\author[b,c]{J. Y. Tang,}
\author[a]{J. D. Liu,}
\author[a]{H. J. Zhang,}
\author[a,1]{B. J. Ye}

\affiliation[a]{State Key Laboratory of Particle Detection and Electronics, University of Science and Technology of China,\\Hefei 230026, China}
\affiliation[b]{Institute of High Energy Physics, Chinese Academy of Sciences (CAS),\\Beijing 10049, China}
\affiliation[c]{Spallation Neutron Source Science Center (SNSSC),\\Dongguan 523803, China}

\emailAdd{panzw19@ustc.edu.cn, bjye@ustc.edu.cn}

\abstract{A new muon source has been proposed to conduct muon spin rotation/relaxation/resonance ($\mu$SR) measurements at China Spallation Neutron Source (CSNS). Only 1 Hz of the CSNS proton beams (25 Hz in total) will be allocated for muon production. To make better use of muons in every pulse, an ultrahigh-array $\mu$SR spectrometer (UHAM) with thousands of detector channels is under design. Due to such a high granularity of detectors, multiple counting events generated from particle scattering or spiral motions of positrons in a strong longitudinal field should be carefully considered in the design. Six different structures were modeled and simulated based on two types of angular orientations (parallel arrangement and pointing arrangement) and three kinds of spectrometer geometries (cylinder, cone and sphere). A quality factor, $Q$, was proposed to evaluate the performance of these structures by integrating their impacts on the overall asymmetry, the counting rate and the percentage of multiple counts. According to the simulation, the conical structure with detectors pointing to the sample has the highest $Q$ in both zero field and longitudinal field. The results also show that all kinds of structures cannot be operated under strong longitudinal fields with a strength over 2 T. The full simulation of a $\mu$SR spectrometer can provide good guidance for the construction of the UHAM in the upcoming upgrade of CSNS.}


\begin{document}
\maketitle
\flushbottom

\section{Introduction}
\label{sec:intro}

Muon is one type of lepton, possessing a spin of 1/2 and a unit electric charge. Muons can be produced in pion decays that are generated from the bombardment of high energetic particles on a production target. In addition, pions decay into muon neutrinos. Due to momentum conservation, muons emit antiparallel to muon neutrinos with their spin antiparallel to momentum. Muons are also unstable and decay into positrons, muon neutrinos and electron neutrinos with a lifetime of $\sim$~2.2 $\upmu$s. According to the parity violation in weak interactions, positrons emit asymmetrically in space and preferentially along the direction of their parental muon spin. The precession of muon spin in magnetic fields makes muons sensitive to the local magnetic property of materials. Therefore, the magnetic properties of materials can be characterized by muon spin rotation/relaxation/resonance ($\mu$SR) techniques.

For general $\mu$SR experiments, a muon beam, a spectrometer and a sample environment are all required in muon facilities~\cite{PSI,TRIUMF,ISIS,J-PARC,MuSIC}. For different experimental purposes, the $\mu$SR spectrometer can work in zero or nonzero magnetic fields. The precision of $\mu$SR data directly depends on the counting rate and the asymmetry of the spectrometer. The counting rate depends on the beam intensity, the solid angle occupied by the detector system, and the pile-up rejection capability of a single detector. The placement of the detector system also determines the asymmetry of the spectrometer. In a pulsed muon beamline~\cite{ISIS,J-PARC,MuSIC}, a detector records all events once their amplitudes are higher than the threshold. It is possible that the multiple counts are generated from one positron in a muon decay event. There are mainly three types of multiple counting events: 1) positrons penetrate multiple adjacent detectors by scattering or spiral motions in a strong magnetic field, 2) secondary particles such as electrons or gamma rays are generated by positrons or muons, and 3) positrons are generated outside the sample. For the first two cases, both positrons and their induced secondaries carry the same asymmetry information. This leads to errors in the counts of each detector, which affects the asymmetry of different detector rings. Consequently, the statistical errors of the data will increase~\cite{Lord2011}. For the last case, the real information of the local field of materials will be mixed with those from outside the sample~\cite{Lynch2003}.

A muon source was proposed in Phase II of China Spallation Neutron Source (CSNS)~\cite{Tang2018, Pan2019, Ni2019, Pan2021, Zhou2022}. Approximately 4\% of the proton beam power (500 kW, 1.6 GeV, 25 Hz) will be assigned to the muon source. Only 1 Hz of the proton repetition rate will be used to make muons. Due to the limit of the repetition rate, an ultrahigh-array $\mu$SR spectrometer (UHAM) dedicated for this muon source should have several thousand detector units to make use of each muon as much as possible. The detector material is made of plastic scintillators with a model type of EJ200~\cite{EJ}. Thus, the total counting rate of the spectrometer is comparable with those in ISIS~\cite{ISIS} and J-PARC~\cite{J-PARC}. However, the percentage of multiple counts will increase due to the high granularity. Hence, the percentage of multiple counts was used as an important parameter to evaluate the performance of a spectrometer in addition to the counting rate and asymmetry. Accordingly, a quality factor, $Q$, was proposed to comprehensively evaluate these parameters.

In the design of the UHAM, there are two options for the arrangement of detectors. Detectors can be placed parallelly along the beam direction or pointing to the sample, which are called the parallel arrangement or pointing arrangement, respectively. In addition, detectors can be placed in various geometries including cylinders, cones and spheres. Considering the two scenarios above, six different detector structures have been modeled. In this work, the quality factors of these structures have been simulated and optimized using musrSim~\cite{Sedlak2012} which is compiling on the Geant4 toolkit~\cite{g1, g2, g3}.

\section{Spectrometer structure}
\label{sec:single}
\begin{figure}[h]
	\centering
	\includegraphics[width=1.0\textwidth]{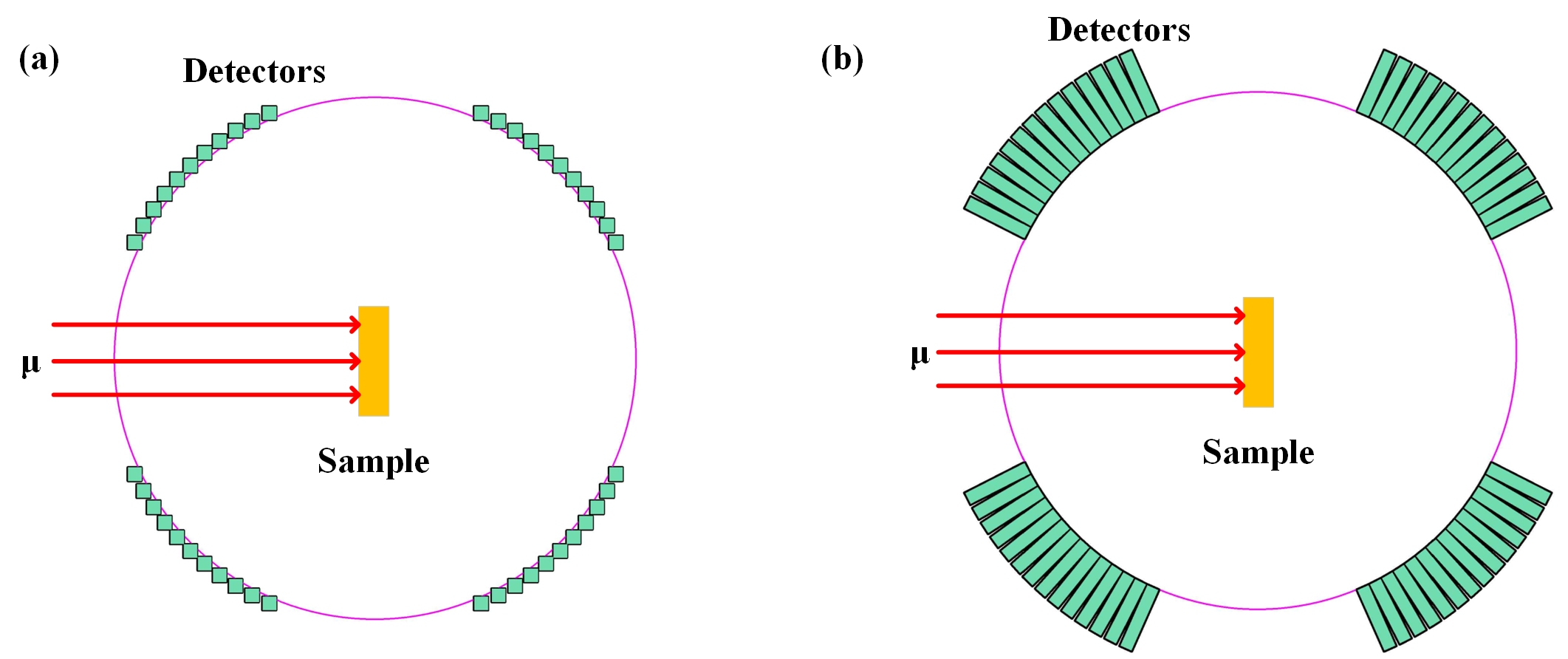}
	\caption{Schematic illustration of (a) parallel and (b) pointing arrangement of detectors in a $\mu$SR spectrometer. Circles in both panels are envelopes of detectors on their inner surfaces.}
	\label{fig1}
\end{figure}

\begin{figure}[h]
	\centering
	\includegraphics[width=1.0\textwidth]{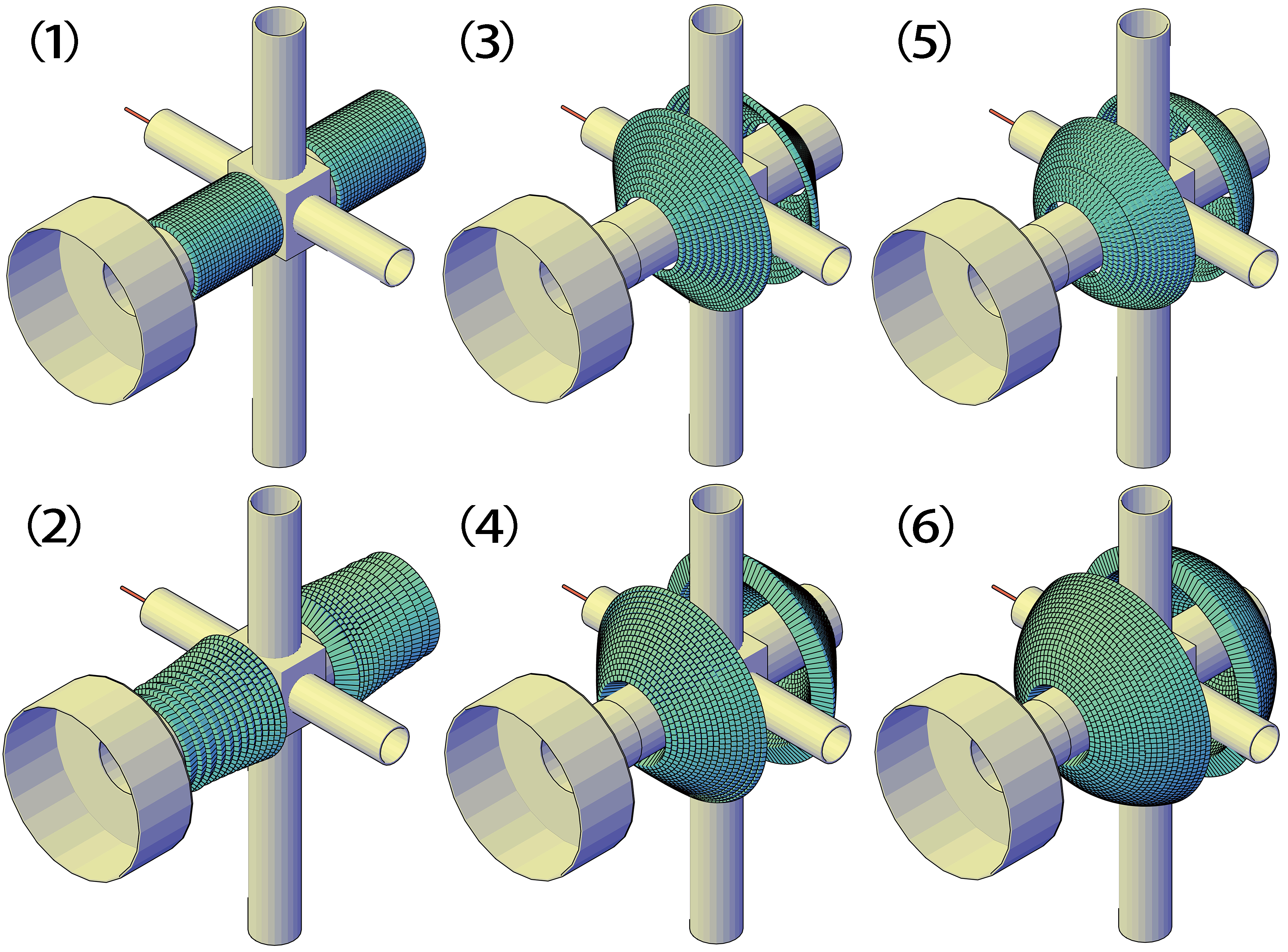}
	\caption{Schematic diagram of the six UHAMs. Detectors in upper panels are placed parallelly, while those in the bottom are placed pointing to the sample. The spectrometer geometries are (1, 2) cylinder, (3, 4) cone, and (5, 6) sphere.}
	\label{fig2}
\end{figure}

\begin{figure}[h]
	\centering
	\includegraphics[width=0.8\textwidth]{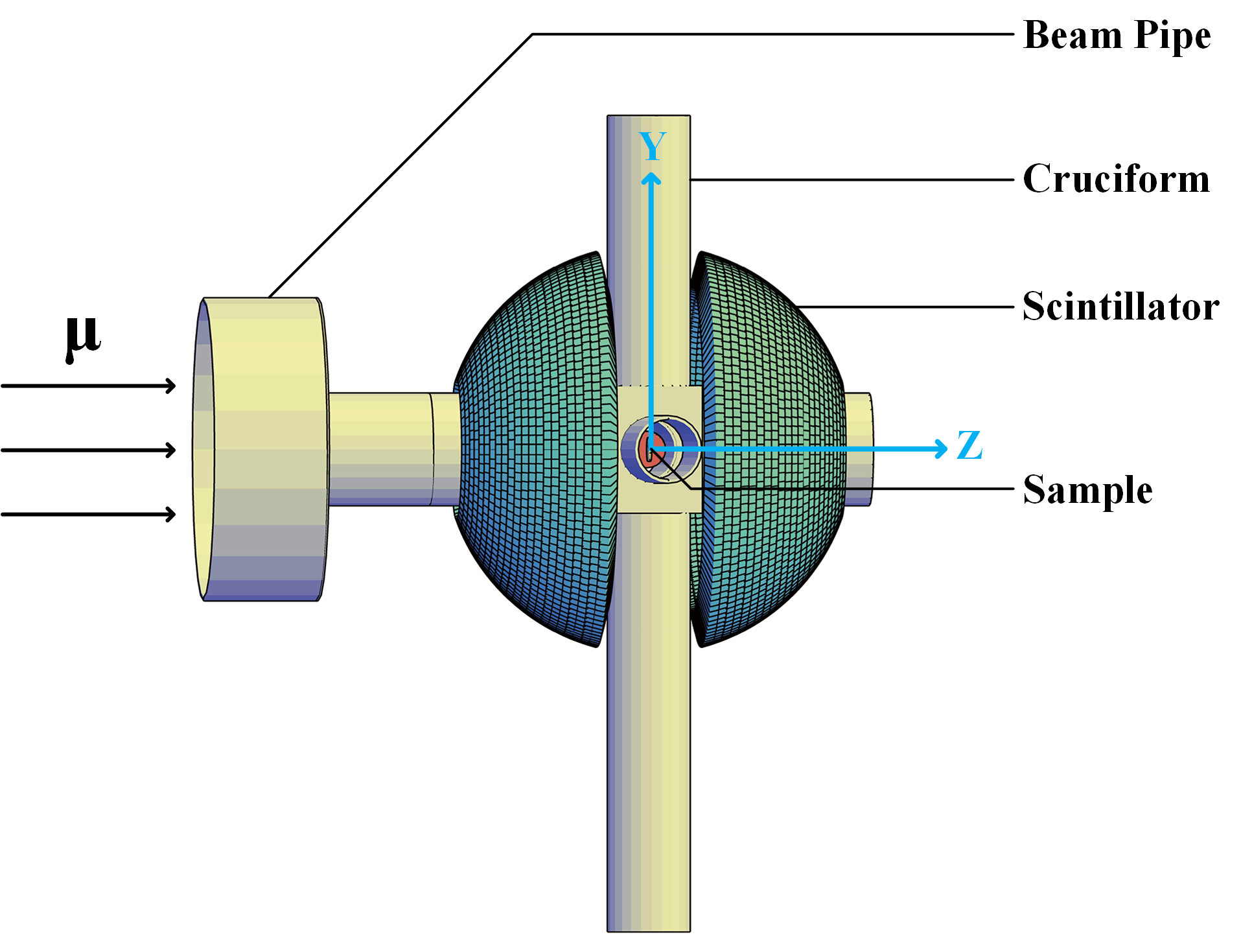}
	\caption{Schematic diagram of spherical geometry with a pointing arrangement.}
	\label{fig3}
\end{figure}

As illustrated in figure~\ref{fig1}, detectors can be placed either parallelly along the beam direction~\cite{Afonso2015, Giblin2014, Salman2009} or pointing to the sample position~\cite{Kadono1996, Tomono2010}. In the parallel arrangement shown in figure~\ref{fig1}(a), detector arrays are placed step by step like stairs. Therefore, every positron decaying from the sample can only penetrate one scintillator along its trajectory. To place several thousand scintillators in a compact geometry, the length (along the beam direction) of each scintillator should be relatively short. In this work, the dimensions of detectors with a parallel arrangement are uniformly set to $10\times 10 \times 10$ mm$^3$. In the pointing arrangement shown in figure~\ref{fig1}(b), the cuboid shape was selected for every single detector instead of the frustum shape to enhance the angle discrimination ability~\cite{Dong2022,Tomono2009}. Different from the parallel arrangement, the length of detectors with a pointing arrangement is not limited, as any positron decaying from the sample only penetrates one scintillator. According to a previous study~\cite{Dong2022}, scintillators with a length of 50 mm are longer enough to achieve the optimal angular discrimination ability. Therefore, the dimension of scintillators with the pointing arrangement is set to $10\times 10 \times 50$ mm$^3$. Additionally, scintillators with dimensions of $10\times 10 \times 10$ mm$^3$ are also placed with the pointing arrangement for the comparison of the parallel arrangement with the same detector sizes. For the geometry of the whole spectrometer, detectors can be placed like cylinders, cones or spheres. Accordingly, six types of UHAMs are simulated in this work, as shown in figure~\ref{fig2}. The detector length of structures ``2'', ``4'', and ``6'' can be either 10 or 50 mm. Each structure is tagged as ``A-B'' in which ``A'' denotes the structure number (1 -- 6 shown in figure~\ref{fig2}), and ``B'' denotes the detector length (10 or 50 mm). For every structure as exemplified in figure~\ref{fig3}, the components modeled in the simulation include a beam pipe, a cruciform-like sample chamber, a silver sample and related sample stick, and plastic scintillators.

As shown in figure~\ref{fig1}, the envelopes of the detectors resemble their spectrometer geometries (cylinder/cone/sphere). For UHAMs with the same envelopes, structures can be approximately set with the same envelope. Therefore, each structure covers the same solid angle with either a parallel arrangement or a pointing arrangement. If so, cylindrical geometry with a pointing arrangement can only place a small number of detectors with large separations between detector rings to avoid geometric overlaps. In contrast, cylindrical geometry with a parallel arrangement can place a large number of detectors as detector separations are not needed. Therefore, the envelope of cylindrical geometries with a parallel arrangement and a pointing arrangement are not the same to compromise the number of detectors and the covered solid angle.

Below are the descriptions of the envelopes for the six spectrometer structures. Table~\ref{tab:Super} lists the number of detector rings, the total number of detectors and the percentage of covered solid angle of the six different UHAMs.

\begin{itemize}
	\item [1)]Structure ``1'': For a cylindrical geometry with a parallel arrangement, its envelope is approximately expressed as:
	\begin{equation}
		x^2+y^2=86^2 ~ (85 {\rm ~ mm}\le |z| \le 335{\rm ~ mm})
		\label{cylinder}
	\end{equation}
	where x, y and z are the position of a scintillator on the envelope in mm. Equation \eqref{cylinder} describes a standard cylindrical surface. This structure has 50 detector rings and 2650 detectors.
	
	\item [2)]Structure ``2'': For a cylindrical geometry with a pointing arrangement, its envelope is approximately expressed as:
	\begin{equation}
		x^2+y^2=96^2 ~ (80 ~ {\rm mm}\le |z| \le 321 ~ {\rm mm})
		\label{cylinder2}
	\end{equation}
	This structure has 26 detector rings and 1586 detectors.
	
	\item [3)]Structures ``3'' and ``4'': For a conical geometry with a parallel arrangement or a pointing arrangement, its envelope is approximately expressed as:
	\begin{equation}
		x^2+y^2=(z-283)^2 ~ (83 ~ {\rm mm}\le |z| \le 200 ~ {\rm mm})
		\label{cone}
	\end{equation}
	 Equation \eqref{cone} describes a standard conical surface. The conical geometry with a parallel arrangement has 26 detector rings and 2468 detectors, while the geometry with a pointing arrangement has 34 detector rings and 2932 detectors.
	
	\item [4)]Structure ``5'' and ``6'': For a spherical geometry with a parallel arrangement or a pointing arrangement, its envelope is approximately expressed as:
	\begin{equation}
		x^2+y^2+z^2=229^2 ~ (65 ~ {\rm mm}\le |z| \le 213 ~ {\rm mm})
		\label{sphere}
	\end{equation}
	Equation \eqref{sphere} describes a standard spherical surface. The spherical geometry with a parallel arrangement has 32 detector rings and 3408 detectors, while the geometry with a pointing arrangement has 44 detector rings and 4290 detectors.
	
\end{itemize}

\begin{table}
	\centering
	\renewcommand\arraystretch{1.0}
	\centering
	\caption{Basic parameters of different UHAMs.}
	\label{tab:Super}
	
	\begin{tabular}{|c|ccccc|}
	\hline	
		Structure ID & Arrangement & Geometry & Detector Ring & Detector & Solid Angle (\%)\\
		\hline
		1 & parallel & cylinder & 50 & 2650 & 30.6   \\
		2 & pointing & cylinder & 26 & 1586 & 36.5   \\
		3 & parallel & cone     & 26 & 2468 & 60.0   \\
		4 & pointing & cone     & 34 & 2932 & 60.4   \\
		5 & parallel & sphere   & 32 & 3408 & 63.2   \\
		6 & pointing & sphere   & 44 & 4290 & 64.9   \\
		\hline
	\end{tabular}
\end{table}

\section{Results and discussion}
\subsection{Simulations in the zero field}\label{sec:zf}

To comprehensively analyze the influence of spectrometer structures on the asymmetry and the counting rate, a figure of merit (FoM)~\cite{Afonso2015} is defined as:
\begin{equation}
	\label{FoM}
	{\rm FoM}=A^2R
\end{equation}
where $A$ is the asymmetry, and $R$ is the counting rate. The counting rate is normally expressed as the number of positrons recorded per unit time. In this work, $R$ is defined as:
\begin{equation}
	R=\frac{N}{N_0}
\end{equation}
where $N$ is the number of positrons counted by detectors, and $N_0$ is the total number of muons. In this work, each simulation generates a number of $10^7$ muons to suppress statistical errors.

Positrons with higher kinetic energy carry higher asymmetries. Placing degraders before detectors to shield low energy positrons is an effective way to enhance the asymmetry of the whole $\mu$SR spectrometer. In this work, brass is used as the degrader material. It can be replaced by other types of materials that have similar shielding effects but with different thicknesses.

\begin{figure}[h]
	\centering
	\includegraphics[width=1.0\textwidth]{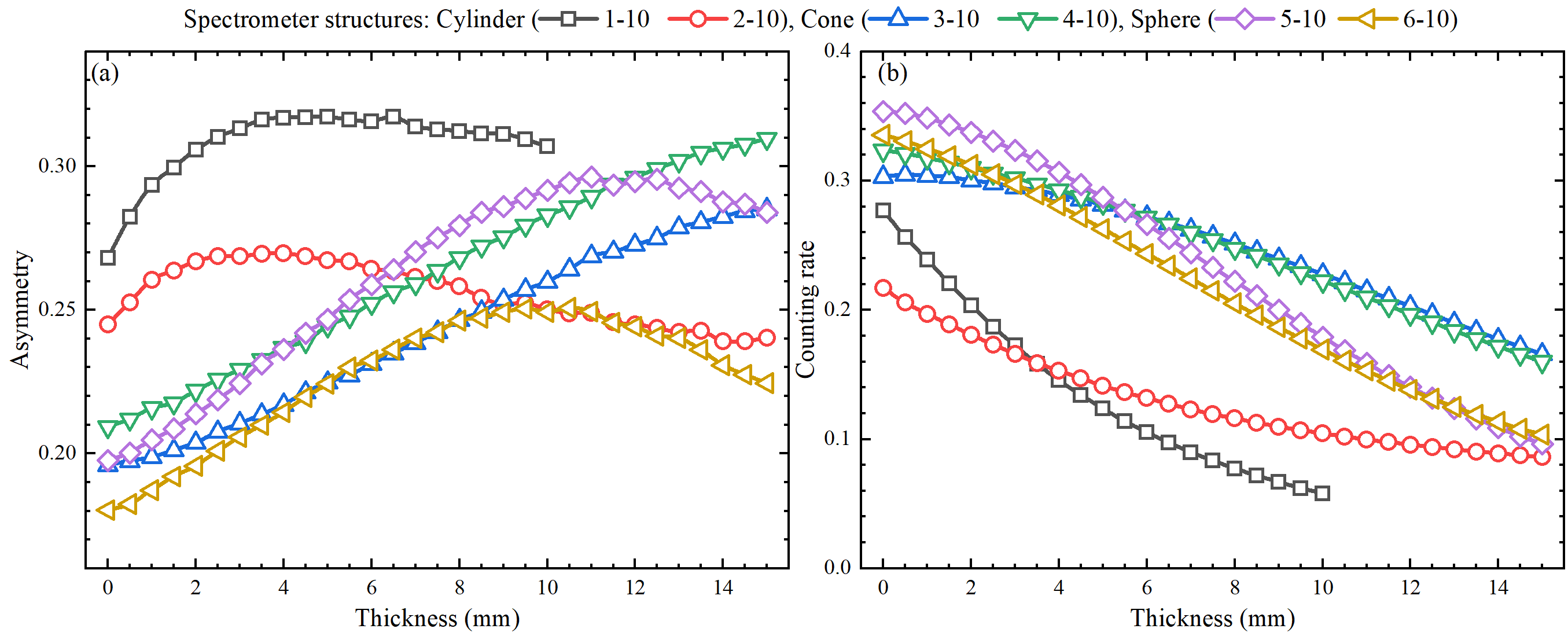}
	\caption{(a) Asymmetry and (b) counting rate of six UHAMs as a function of the degrader thickness. Curves are visual guides.}
	\label{fig4}
\end{figure}

The length of detectors for all spectrometer structures is set to 10 mm in the beginning to take a quick look at the impacts of degraders on the counting rate and the asymmetry. It should be noted that the degrader thickness of structure ``1-10'' can only reach 10 mm, otherwise geometric overlap will occur. As shown in figure~\ref{fig4}(a), the asymmetry of all structures gradually increases to the maximum and then decreases as a function of the degrader thickness. Structures with cylindrical geometries reach the maximum asymmetry more quickly followed by spherical and conical geometries. For structures with the same geometry, detectors placed parallel to the beam direction achieve higher asymmetries. In contrast, the counting rate of all structures decreases as a function of the degrader thickness as shown in figure~\ref{fig4}(b). When the degrader thickness is zero, the differences in the counting rates of the three spectrometer geometries are consistent with the differences in their covered solid angles as listed in table~\ref{tab:Super}. For any given spectrometer geometry, detectors with a parallel arrangement have a higher counting rate than those with a pointing arrangement (for example, structure ``1-10'' > ``2-10''). Such an effect (a lower covered angle leads to a higher counting rate) is related to the multiple counting which will be discussed later. The rank of the counting rate changes when increasing the degrader thickness particularly with sizes over $\sim$~5 mm.

\begin{figure}[h]
	\centering
	\includegraphics[width=0.5\textwidth]{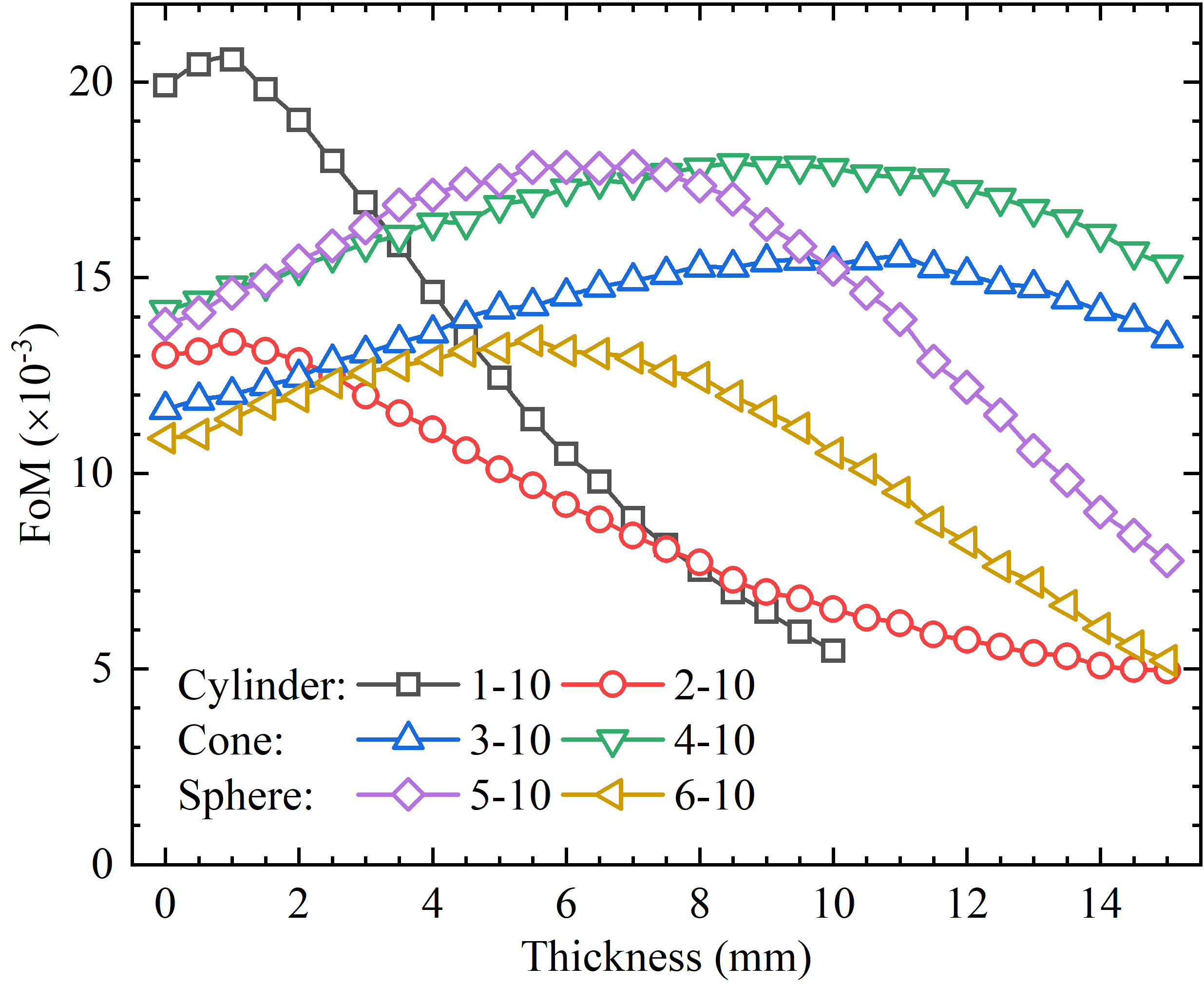}
	\caption{FoM of different UHAMs as a function of the degrader thickness. Curves are visual guides.}
	\label{fig5}
\end{figure}

Figure~\ref{fig5} presents the FoMs of six UHAMs based on equation~\eqref{FoM}. The variation trend of FoM behaves similarly to that of the asymmetry (see figure~\ref{fig4}(a)) as a function of the degrader thickness. The cylindrical geometry rapidly peaks at a degrader thickness of $\sim$~1 mm, while the spherical and conical geometries gradually reach their maximum asymmetries at $\sim$~6 mm and $\sim$~10 mm, respectively. For cylindrical and spherical geometries, detectors with a parallel arrangement have higher asymmetries than those with a pointing arrangement. This difference is inverse for the conical geometry.

\begin{figure}[h]
	\centering
	\includegraphics[width=1.0\textwidth]{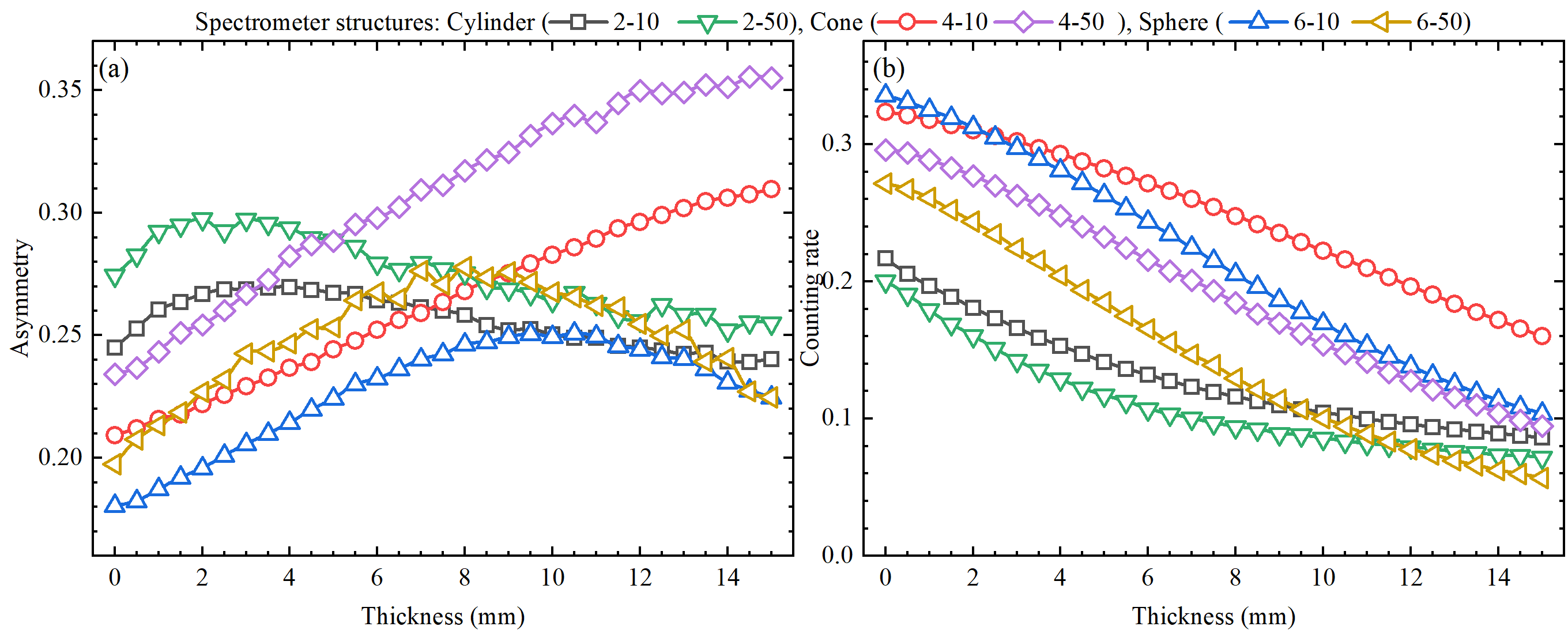}
	\caption{(a) Asymmetry and (b) counting rate of UHAMs with pointing arrangements and different detector lengths. Curves are visual guides.}
	\label{fig6}
\end{figure}

\begin{figure}[h]
	\centering
	\includegraphics[width=0.5\textwidth]{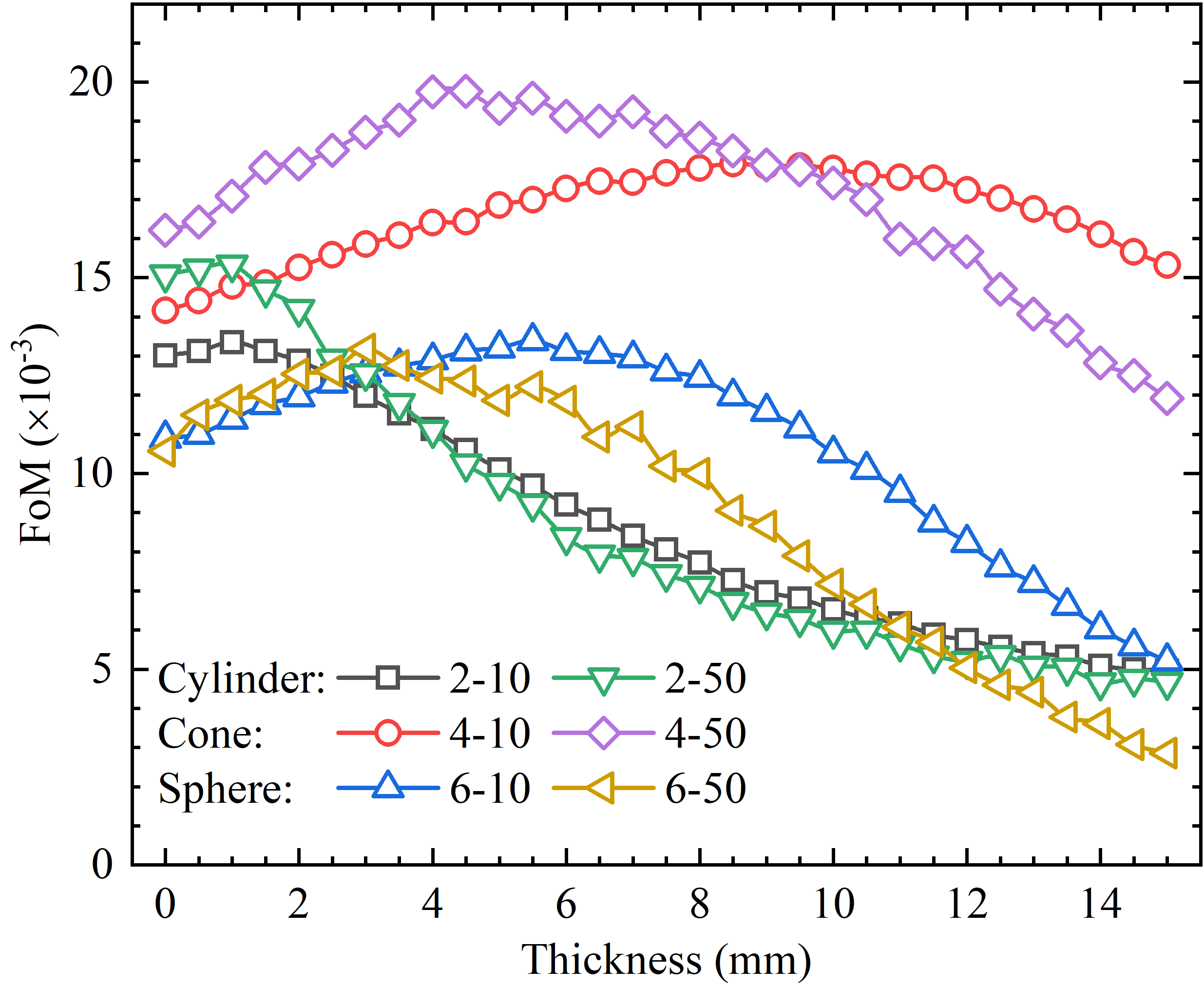}
	\caption{FoMs of different UHAMs with pointing arrangements and different detector lengths. Curves are visual guides.}
	\label{fig7}
\end{figure}

The comparison of UHAMs with different detector lengths is presented in figure~\ref{fig6} for the pointing arrangement. The variation trends of the asymmetry or the counting rate of spectrometers with different detector lengths are almost the same for any given geometry. However, detectors with a length of 50 mm have higher asymmetries but lower counting rates than those with a length of 10 mm for the same geometry. This is due to the energy threshold setup. The mean energy deposition of positrons in 50-mm detectors is nearly 4 times higher than that in 10-mm ones. A much higher threshold is set for the longer detectors. As a result, the 50-mm detectors lose some percentage of positrons. Positrons with higher energy carry higher asymmetries. A higher threshold for longer detectors leads to higher asymmetries for the whole spectrometer. Figure~\ref{fig7} gives the FoMs of these spectrometer structures, which present similar increasing-decreasing trends. The maximum FoMs for 50-mm detectors are slightly higher than those for 10-mm detectors for any given spectrometer geometry. Among them, the conical geometry with a 50-mm detector achieves the highest FoM.

\begin{figure}[h]
	\centering
	\includegraphics[width=0.5\textwidth]{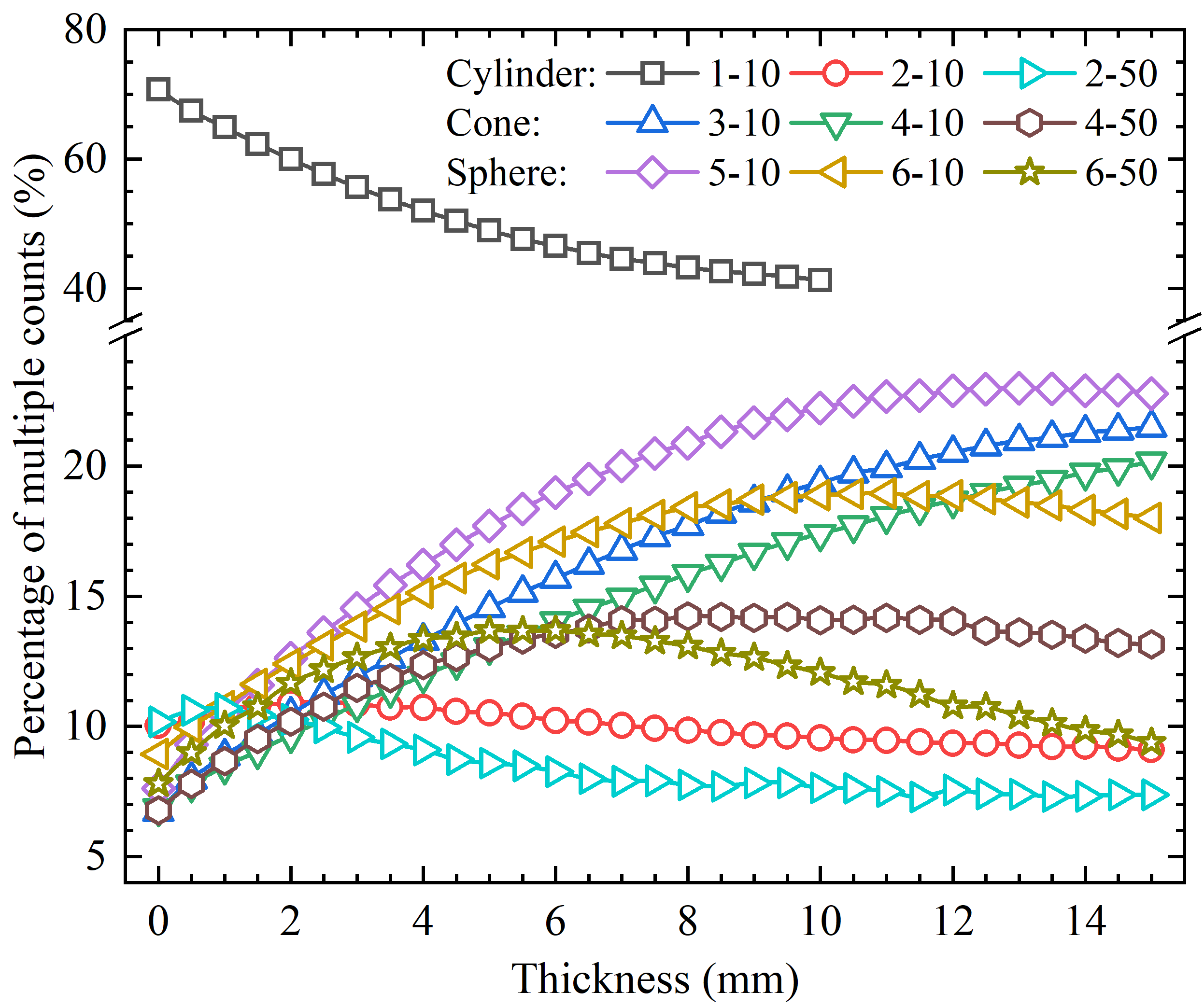}
	\caption{Percentage of multiple counts of different UHAMs as a function of the degrader thickness. Curves are visual guides.}
	\label{fig8}
\end{figure}

As mentioned in section~\ref{sec:intro}, the performance of a $\mu$SR spectrometer can be affected by the multiple counting effects with high granularity. According to the running experience of HiFi at ISIS Muon Facility, the $M$ is tolerable below 30\%~\cite{Lord2011, Afonso2015, Salman2009}. Figure~\ref{fig8} presents the percentage of multiple counts (denoted as ``$M$'') of all 9 spectrometer structures simulated above. It is obvious that structure ``1-10'' has the highest $M$. Due to the placement of scintillators in detector rings with a fixed radius along the beam direction, positrons can easily pass through multiple adjacent detectors along their trajectories at the same time. In addition, the variation trend of structure ``1-10'' is different from all other structures as a function of the degrader thickness. The $M$ for other structures increases to the maximum and then has a decreasing trend. Comparing detectors with the same length (``1-10'' with ``2-10'', ``3-10'' with ``4-10' and ``5-10'' with ''6-10'') and the same geometry, structures with a pointing arrangement obtain lower $M$. This is due to the relatively better angular discrimination ability with pointing arrangements~\cite{Dong2022}. According to our previous study~\cite{Dong2022}, the angular discrimination ability reaches the maximum when the detector length is over $\sim$~50 mm. Therefore, multiple counts are significantly suppressed in pointing arrangements with a detector length of 50 mm.

\begin{figure}[h]
	\centering
	\includegraphics[width=0.5\textwidth]{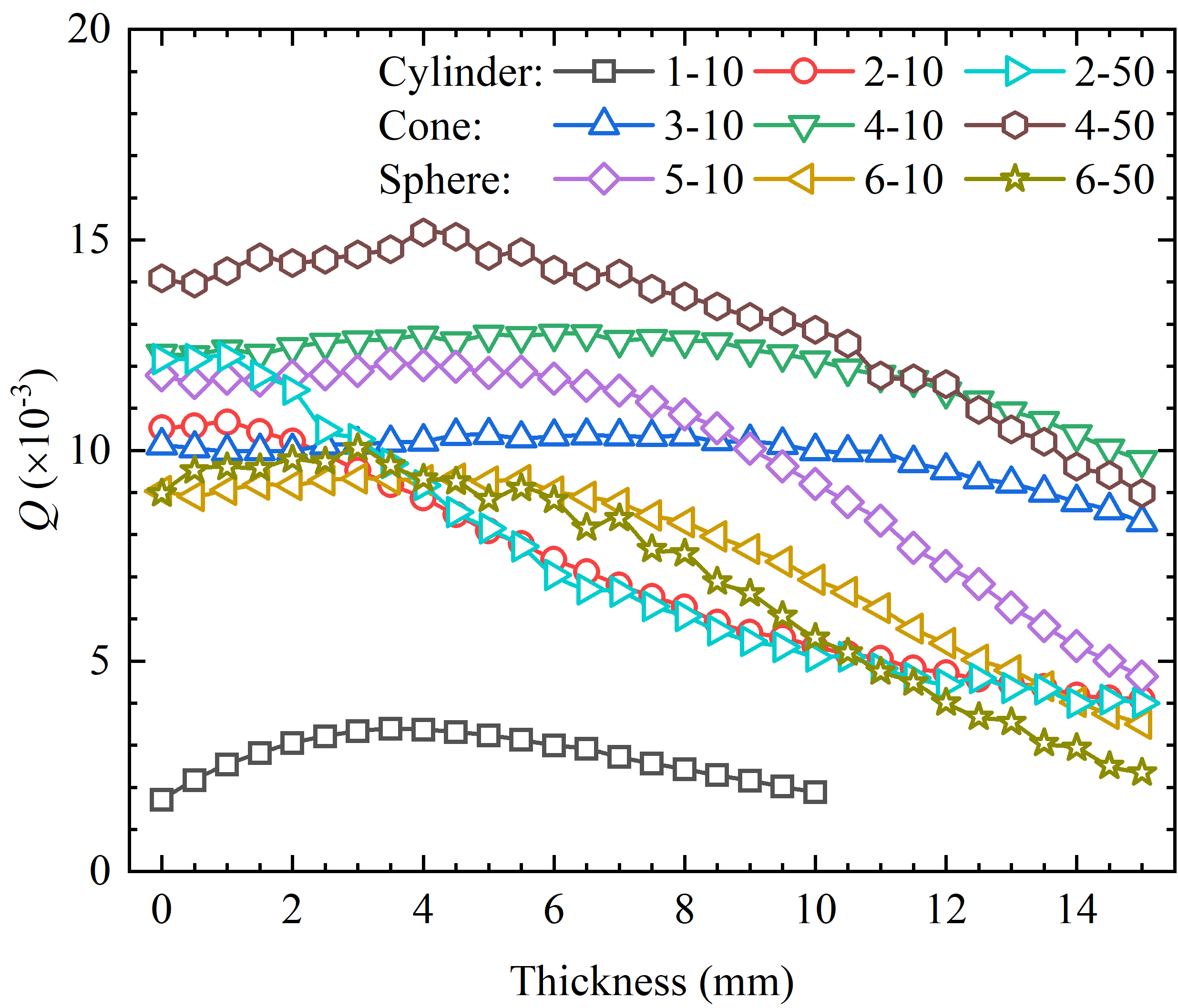}
	\caption{Quality factors of different UHAMs in variation with degrader thickness. Curves are visual guides.}
	\label{fig9}
\end{figure}

To comprehensively consider the influences of spectrometer structures on $A$, $R$ and $M$, a quality factor, $Q$, can be used as:
\begin{equation}
	Q= A^2R(1-M)^2
	\label{eq-Q}
\end{equation}
Accordingly, the $Q$ values for all mentioned structures are presented in figure~\ref{fig9}. All data show an increasing-decreasing trend as a function of the degrader thickness. It is obvious that the structure ``4-50'' achieves the optimal $Q$ when the degrader thickness is $\sim$~4 mm. Table~\ref{tabQ} lists the degrader thickness, $M$, $A$, and $R$ for all structures at their maximum $Q$ values.

\begin{table}[h]
	\renewcommand\arraystretch{1.0}
	\centering
	\caption{Maximum $Q$ and related parameters of different UHAMs in the zero field.} 
	\label{tabQ}
	\begin{tabular}{|c|ccccc|}
		\hline
		Structure ID & Maximum $Q$($\times 10^{-3}$) & Degrader Thickness (mm) & $M$ (\%) & $A$ & $R$\\
		\hline
		1-10    & 3.38 & 3.5 & 53.8 & 0.316 & 0.158 \\
		2-10    & 10.7 & 1.0 & 10.6 & 0.261 & 0.197 \\
		3-10    & 10.4 & 5.0 & 14.5 & 0.225 & 0.281 \\
		4-10    & 12.8 & 5.0 & 13.0 & 0.244 & 0.282 \\
		5-10    & 12.1 & 3.5 & 15.4 & 0.231 & 0.315 \\
		6-10    & 9.33 & 3.0 & 13.8 & 0.206 & 0.297 \\
		2-50    & 12.2 & 1.0 & 10.7 & 0.292 & 0.179 \\
		4-50    & 15.2 & 4.0 & 12.4 & 0.282 & 0.248 \\
		6-50    & 10.1 & 3.0 & 12.6 & 0.243 & 0.224 \\
		\hline
	\end{tabular}
\end{table}

\subsection{Simulations in longitudinal fields}\label{sec:lf}
Apart from the zero field, magnets that can generate a tunable longitudinal field are also essentially required in $\mu$SR spectroscopy. To measure the full muonium repolarization curve in a longitudinal field, the field strength should have the power to be scanned from 0 to up to 5000 G~\cite{Nuccio2014}. In addition, a $\mu$SR spectrometer that operates with high magnetic fields in the few Tesla range enables the investigation of a wider range of dynamics, access to different regions in the phase diagram, spectroscopy and molecular dynamics of various new systems and the ability to perform a combination of RF-$\mu$SR and NMR experiments~\cite{Salman2009, King2003}. Therefore, two longitudinal field ranges (0 - 5000 G, and 0 - several Tesla) are normally deployed in international muon facilities~\cite{ISIS-I, PSI-I, TRIUMF-I}. In Phase II of CSNS, the construction of the first $\mu$SR spectrometer will start with a longitudinal field lower than 5000 G. In this work, a longitudinal field range of (0, 10 T) is scanned in simulations to study the feasible field range of the current spectrometer design.

\begin{figure}[h]
	\centering
	\includegraphics[width=1.0\hsize]{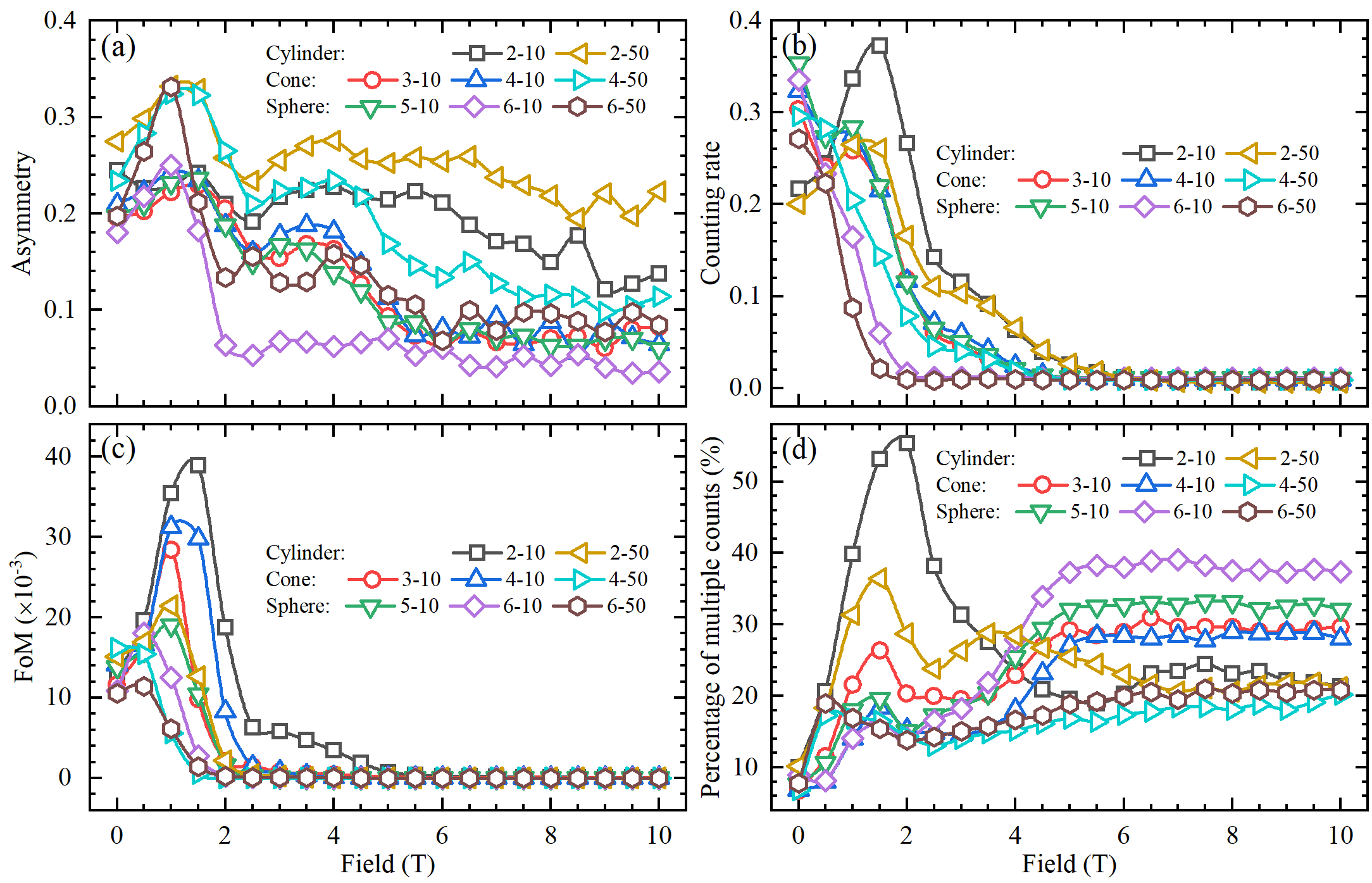}
	\caption{(a) The asymmetry, (b) counting rate, (c) FoM and (d) percentage of multiple counts as a function of the field strength. Curves are visual guides.}
	\label{fig10}
\end{figure}

\begin{figure}[h]
	\centering
	\includegraphics[width=0.5\textwidth]{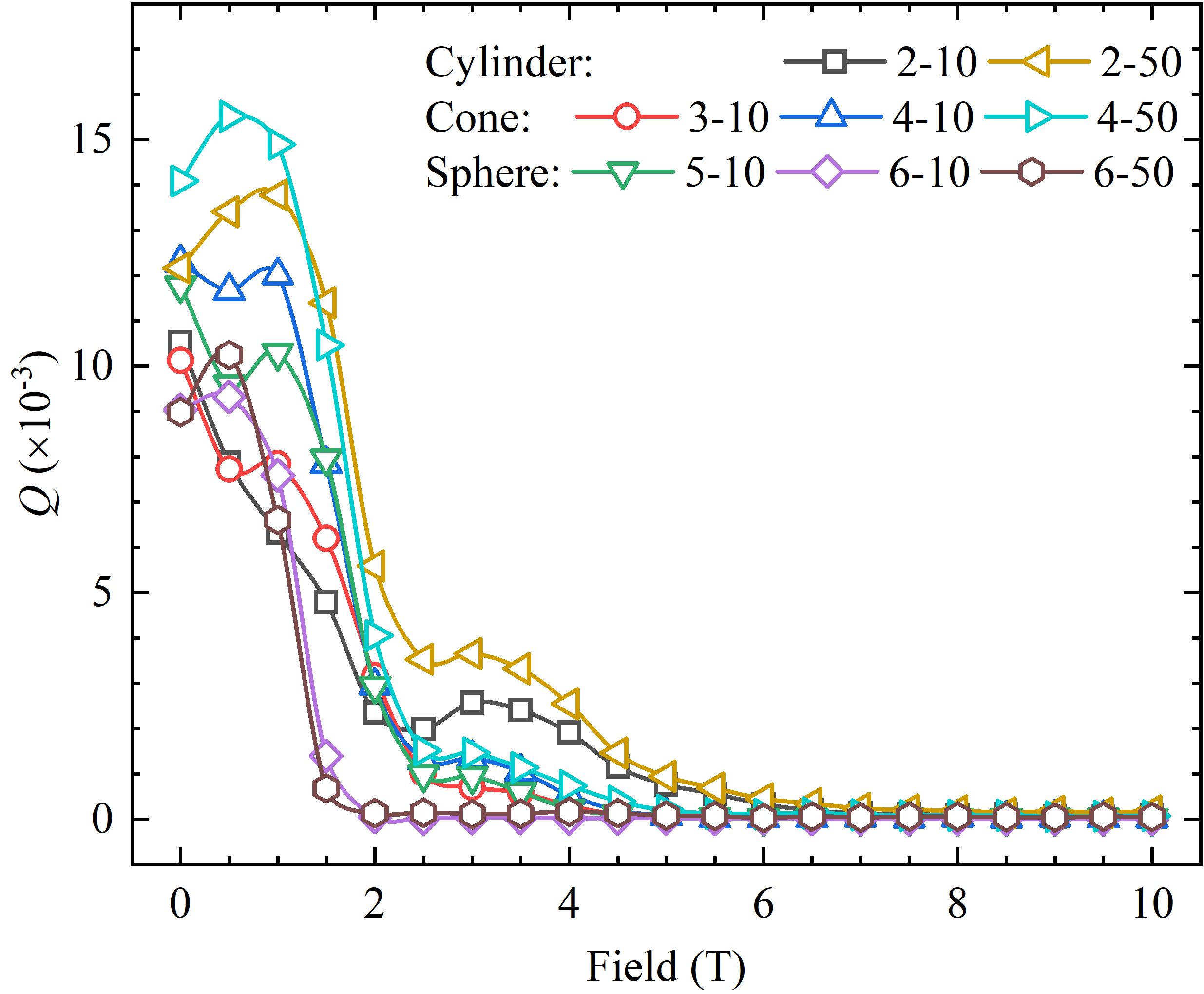}
	\caption{Quality factor of different UHAMs in variation with magnetic field. Curves are visual guides.}
	\label{fig11}
\end{figure}

As the $M$ of structure ``1-10'' is quite high in the zero field, it is not considered in the longitudinal field simulations. Figure~\ref{fig10} presents the $A$, $R$, FoM and $M$ of different UHAMs scanning the longitudinal field from 0 to 10 T with a step length of 0.5 T. The asymmetry in figure~\ref{fig10}(a) peaks at $\sim$~1 T and then decreases with some fluctuations. The variation trend of the counting rate in figure~\ref{fig10}(b) is different between the cylindrical geometry and the other two geometries. The counting rate of the cylindrical geometry peaks at a field strength of $\sim$~1.5 T, while the other geometries continuously reduce to 0. As positrons do spiral motions in a strong longitudinal field, the radius of spiral motions is inversely proportional to the field strength. Positrons cannot hit detectors with the radius of detector rings larger than the radius of spiral motions. Hence, the counting loss of detectors tends to be more obvious with conical and spherical geometries in strong fields. According to the variation trends of the asymmetry and the counting rate, the FoM in figure~\ref{fig10}(c) peaks at a field strength followed by rapid reduction. The percentage of multiple counts in figure~\ref{fig10}(d) confirms that the spiral motions of positrons lead to an increase in multiple counting probability. For any given spectrometer geometry, detector placed pointing to the sample with a length of 50 mm can more easily suppress multiple counting mainly due to the spiral motions of positrons. Figure~\ref{fig11} presents the overall influences integrating all parameters of figure~\ref{fig10}. In the range of (0, 2 T), the structure ``4-50'' has the optimal $Q$ over the other ones. For the field strength over 2 T, the $Q$ of all structures reduces to very low values compared with their initial values in the zero field. Therefore, the design of UHAMs in this work can be well performed in the required longitudinal field (< 5000 G). The design accepts the upgrade of the longitudinal field up to 2 T in the future. If a field over 2 T is needed, the arrangement of the UHAM should be updated to fit the strong field.

According to the simulation in the zero field and a tunable longitudinal field, the conical geometry with a pointing arrangement and a detector length of 50 mm has the optimal quality factors. This can be a reasonable reference for the construction of the first $\mu$SR spectrometer at CSNS.

\section{Conclusions}\label{sec:conclu}
In this work, six different structures of UHAMs have been modeled and simulated for the construction of a muon source for CSNS. According to the requirement of a magnetic field in real $\mu$SR experiments, the simulation is performed in both the zero field and a tunable longitudinal field. The performance of these structures are compared by their overall asymmetries, counting rate and percentage of multiple counts. A quality factor is proposed to comprehensively compare these parameters. The simulation confirms that detectors pointing to the sample have the advantage of multiple counting suppression. Detectors with a large aspect ratio can enhance this capability. Simulations in both the zero field and longitudinal field support that the spectrometer structure has optimal performance with a conical geometry, a pointing arrangement and a detector length of 50 mm. For longitudinal field experiments, the spectrometer design can work well in the required field range (< 5000 G). It has the potential to be operated up to 2 T if the magnets are required to be updated to extend the research field of $\mu$SR techniques at CSNS.

\acknowledgments
This work was financially supported by the National Natural Science Foundation of China under Grants 12005221 and 11527811. The authors are grateful to the accelerator group of CSNS for their helpful discussions and comments.

\end{document}